\documentclass[conference]{IEEEtran}
\ifCLASSINFOpdf
  \usepackage[pdftex]{graphicx}
\else
\fi
\usepackage{balance}

\usepackage{amsmath}
\begin{document}
\title{Urban Heat Islands: Beating the Heat with Multi-Modal Spatial Analysis}

\author{\IEEEauthorblockN{Marcus Yong}
\IEEEauthorblockA{Information Systems Technology and Design Pillar\\
Singapore University of Technology and Design\\
Email: marcus\_yong@mymail.sutd.edu.sg}
\and
\IEEEauthorblockN{Kwan Hui Lim}
\IEEEauthorblockA{Information Systems Technology and Design Pillar\\
Singapore University of Technology and Design\\
Email: kwanhui\_lim@sutd.edu.sg}}


%Telephone: (800) 555--1212, Fax: (888) 555--1212}


% for IEEE copyright statement
% 978-1-7281-6251-5/20/$31.00 ©2020 IEEE
\IEEEoverridecommandlockouts
\IEEEpubid{\makebox[\columnwidth]{978-1-7281-6251-5/20/\$31.00~\copyright~2020 IEEE \hfill} \hspace{\columnsep}\makebox[\columnwidth]{ }}

\maketitle

% for IEEE copyright statement
\IEEEpubidadjcol

% for page number
% \thispagestyle{plain}
% \pagestyle{plain}

\begin{abstract}
In today's highly urbanized environment, the Urban Heat Island (UHI) phenomenon is increasingly prevalent where surface temperatures in urbanized areas are found to be much higher than surrounding rural areas. Excessive levels of heat stress leads to problems at various levels, ranging from the individual to the world. At the individual level, UHI could lead to the human body being unable to cope and break-down in terms of core functions. At the world level, UHI potentially contributes to global warming and adversely affects the environment. Using a multi-modal dataset comprising remote sensory imagery, geo-spatial data and population data, we proposed a framework for investigating how UHI is affected by a city’s urban form characteristics through the use of statistical modelling. Using Singapore as a case study, we demonstrate the usefulness of this framework and discuss our main findings in understanding the effects of UHI and urban form characteristics. 
\end{abstract}


% \fi
\IEEEpeerreviewmaketitle

\section{Introduction}
Heat originates from a myriad of different sources: fossil fuel combustion, vehicular and air-conditioning emissions, commercial and industrial outputs, and human activities, not forsaking metabolism. The emanation and trapping of such heat in an urban concrete jungle leads to the Urban Heat Island (UHI) phenomenon, where surface temperatures are found to be much higher than surrounding rural areas~\cite{akbari1990}. With its dense high-rise buildings, the Metropolis of Singapore is highly susceptible to the Urban Heat Island (UHI) phenomenon. 

Experts have cautioned that Singapore is heating up at twice the rate as the world itself, which is especially worrying as it will significantly increase heat stress when combined with the high humidity levels on the island~\cite{TanTang2019}. Besieged by excessive levels of heat stress, the human body would potentially find difficulty in coping and breakdown~\cite{TanTang2019}. A whole slew of other problems could emerge as well since rising surface temperatures would lead to increased energy consumption for cooling, which contributes to thermal output that feeds the UHI effect spiral. Warmer waters would damage aquatic ecosystems and we can also expect to feed the global warming problem~\cite{NatGeo2011}. Fortunately, studies have found that clever urban design can have a huge impact in mitigating the UHI effect~\cite{Taslim2015,soltanifard2019}. 

This research aims to provide insights into the UHI effect and how it is affected by a city’s urban form characteristics through the use of statistical modelling.  
Using a case study on the city of Singapore, we seek to answer the following research questions:

\begin{itemize}
\item RQ 1: How is the UHI effect influenced by the urban form?
\item RQ 2: Which elements of the urban form are worth investigating?
\end{itemize}

Key characteristics of the urban form are represented by surface and building information that can both be obtained from a diversity of different sources. We employed a mixture of methods that includes the use of readily available remote sensing imagery and geospatial scraping to procure the necessary data for analysis. Although some of the previous research has incorporated 3-dimensional building information data, limited access to such data in Singapore has made it infeasible for our approach. Nonetheless, it is compensated for by employing the use of open-source HDB property information from data.gov.sg that provides useful alternative 2-dimensional data. 2-dimensional data has been proven to be adequate in accurately estimating Land Surface Temperatures~\cite{Carlson1981,Nichol1996} and we adopt similar data sources for our approach.

The rest of this research paper will contain the following sections: Section~\ref{sectLitReview} contains a  literature review of the existing work that has been done, Sections~\ref{sectFramework},~\ref{sectFramework-1} and~\ref{sectFramework-2} consist of the research framework we devised, Section~\ref{sectResults} presents the experimental results and associated discussions, while Section~\ref{sectConclusion} concludes the paper and details areas for future work.

\section{Literature Review}
\label{sectLitReview}

\subsection{Background of the Urban Heat Island effect}

\begin{figure}[h!]
\includegraphics[width=0.4\textwidth]{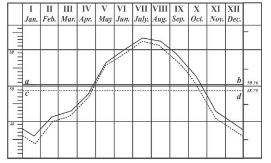}
\centering  
\caption{Comparison of Air Temperatures in 1833: Central London \& Countryside. Retrieved from~\cite{Howard1833}.}
\label{table:map1}
\end{figure}

The Urban Heat Island (UHI) effect is a growing environmental concern that has arisen from an increasingly urbanized world. It was first understood by Luke Howard as the difference in air temperature between the general country of London and that of the Royal Society in central London in 1833. In Figure~\ref{table:map1}, he investigated the difference in air temperatures over a period of a year and his results indicate a higher mean temperature in the city (solid line) as opposed to the countryside (dashed). Howard then concluded that the mean temperature of London’s climate is approximately 48.50 degrees fahrenheit while central London’s mean temperature is higher at 50.50 degrees fahrenheit~\cite{Howard1833}. Luke Howard’s attempt in studying air temperature differences in London laid the groundwork for a growing field of research in UHI, which has been ongoing for many decades~\cite{Oke1973,Oke1981,Chun2014,Zhang2017,BourbiaAwbi2004} 
with increasing levels of complexity. Today, the UHI effect is similarly coined by the phenomena where urban-dense cities have higher temperatures than their surroundings. This is due to how impervious urban surfaces such as concrete and tarmac absorb and contain solar heat energy and release it into the environment to create a warmer microclimate~\cite{Wilhelmi2004}.

In recent years, Tomlinson et al. suggested that the traditional method of Urban Heat Island assessment is often carried out with inaccessible and inadequate meteorology data from weather stations~\cite{Tomlinson2012}. As such, Tsou et al. attempted to overcome these challenges by quantifying the UHI phenomenon using Landsat 8 data through the estimation of Land Surface Temperature (LST) for the urban-dense cities of Shenzhen and Hong Kong~\cite{Tsou2017}. Their study echoed Rao’s idea to deploy remote sensing in the field of UHI studies, which is achievable through the generation of a land surface thermal distribution graph by processing thermal infrared data from satellite images~\cite{Rao1972}.

\subsection{Surface Information (NDVI, NDWI, Population)}
UHI is dependent on a myriad of urban form factors, one of which is the amount of vegetation in the urban environment~\cite{doick2013}. This can be represented by the Normalized Difference Vegetation Index (NDVI) in the form of surface data. Research conducted by Solecki et al.~\cite{solecki2005} suggests that NDVI can be computed via the same remote sensing method as LST computation, and results indicate that NDVI has a negative relationship with UHI. In another study done on UHI mitigation, UHI intensity has been proven to decrease with increasing vegetation~\cite{Aboelata2017}. This can be explained by how vegetation lowers the temperature of the urban environment through evaporative cooling in plants~\cite{Oke2017} while also significantly reducing heat in street canyons. 
Moreover, by studying how vegetation affects UHI in densely-built cities like New York, temperature differences of as much as 2 degrees celsius were observed between areas with the most and the least amount of vegetation in the city~\cite{Susca2011}. 

Similar to vegetation, water bodies are found to have a cooling effect on LSTs in the urban setting. \cite{EstoqueMurayama2017} posits the computation of the Normalized Difference Water Index (NDWI) using surface data to measure water content in an area using the same aforementioned remote sensing method and results reveal a negative relationship with UHI (measured by LST). In the city of Suzhou, water bodies such as Suzhou Bay can effectively reduce temperatures by as much as 3 degrees Celsius at a reach of 800m away~\cite{Wu2019}. Furthermore, the results indicate that the presence of water bodies also boosts the cooling effects of vegetation on LSTs, thereby further enhancing the cooling abilities of vegetation on top of reducing temperatures themselves. This suggests that water bodies should ideally be located together with vegetation in an ecosystem to maximise urban cooling.

With regards to surface population information, another research by Zhang and Wu demonstrates that mass human movement out of an urban area has a significant impact on the temperature of the urban environment~\cite{Zhang2017}. By studying the temporary migration of Chinese during the Chinese New Year holidays out of Beijing, they discovered a statistically significant decline in urban temperature in the absence of the masses in the city, suggesting that population density has significant influence over the UHI effect. 

\subsection{Building Information (Height, Commercial/Industrial Facilities)}
UHI is also influenced by the height of buildings. Building height affects air circulation and thermal energy build-up~\cite{Oke1981,Eliasson1996,BottyanUnger2003}. In spite of how tall buildings are able to reduce surface area exposure to solar radiation by blocking out sunlight and providing shade, research has found that surface temperatures in built-up environments are still relatively higher than rural areas~\cite{ValssonBharat2009}. Work carried out by Giridharan et al.~\cite{Giridharan2004} in densely-packed Hong Kong illustrates that regions with closely-clustered and taller buildings have approximately 1.5 degree higher temperatures than other regions. Cities with high urban density do not release thermal energy into the environment easily and the obstruction of air circulation prevents effective cooling. Solar thermal energy is also retained in the walls of urban structures that further drive up the surrounding temperature. 

UHI is influenced by the presence of commercial and industrial facilities as well. In a study of land use in Singapore and its impact on the UHI effect, Jusuf et al. proved via qualitative and quantitative methods that surface temperatures were highest on land with industrial, commercial and airport facilities, and lowest on land with residential and park facilities during the daytime~\cite{Jusuf2007}. 

\subsection{Statistical Modelling}
The field of UHI research has seen a number of quantitative analytical methods, one of which is a statistical approach of analyzing the UHI effect carried out by Chun and Guldmann. In this study, OLS regression and spatial regression techniques were employed to understand the relationship between UHI, represented by LST, and urban characteristics over different hexagonal grid sizes~\cite{Chun2014}. 
Data describing the urban form such as the vegetation index and building height were collected and factored into the analysis for different parameters of grid structures. Model selection criteria were employed to select the most accurate regression model that describes the relationship between LST and urban form for an optimal grid structure. In another study done by Yin et al. on the city of Wuhan, UHI was similarly spatially analyzed and results indicated that UHI can be mitigated by optimizing urban form along with balancing various land use types~\cite{Yin2018}.

\subsection{Discussion of Existing Literature}
The existing work on UHI has revealed useful variables and methods to consider in the analysis of UHI but it also exposes several areas that could benefit from additional work. They lack the consideration of anthropogenic heat with specific regards to the land use and the archetype of buildings or areas of residence. Furthermore, the research done on UHI is primarily focused on cities where a rural backyard of comparable size is available to be juxtaposed against the metropolitan city centre. Although there has been UHI related work conducted for other cities~\cite{hart2009,soltanifard2019}, little work is done for the unique case of urban-dense Singapore where there is absence of a rural hinterland.

\section{Overview of Research Framework}
\label{sectFramework}

In order to understand how urban form characteristics influence the UHI effect, the research framework is broken down into two main sections, which we describe next.

\subsection{Data Preparation}
The first part of the framework looks at extracting data from multiple sources and combining them into a single, large dataset. Urban form characteristics are represented by variables that are defined at a common geospatial hexagonal grid level (Section~\ref{sectModelSelection}). These urban form variables are categorized into dependent variables and independent variables (surface information, building information) as represented by Table~\ref{depIndepVar}.

\begin{table}[h!]
\includegraphics[width=0.5\textwidth]{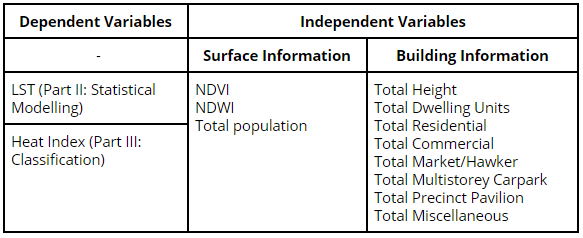}
\centering  
\caption{Overview of all dependent and independent variables}
\label{depIndepVar}
\end{table}

\subsection{Statistical Modelling}
The second part looks at the use of Ordinary Least Squared regression (Section~\ref{sectOLSRegression}) and spatial regression statistical modelling techniques (Sections~\ref{sectSpatialRegression} and~\ref{sectSEM}). We seek to understand how urban form characteristics, consisting of surface and building information, are statistically significant in explaining the UHI effect (represented by LST).

\section{Data Sources and Data Preparation}
\label{sectFramework-1}

The proposed approach to studying the Urban Heat Island effect requires a variety of data and information sources, which is not available in any single dataset. As such, we curated our own dataset from a cumulation of multiple data sources with their own specific type of data, all of which are obtained from online sources:
\begin{itemize}
\item Remote Sensing Dataset (https://earthexplorer.usgs.gov/):
The data comes in a raster file captured by the Landsat 8 satellite at 100m resolutions with the least cloud cover (less than 10\%) on May 24, 2018. It has information about the Earth’s surface that allows us to compute variables such as LST, NDVI, NDWI. 
\item  Residential HDB Building Dataset + (x,y) coordinates (https://data.gov.sg/, https://docs.onemap.sg/):
This dataset provides us with building information of specific HDB flats in Singapore such as the HDB block’s height, presence of commercial facilities, total number of dwelling units, and presence of amenities. The relevant (x,y) coordinates had to be scraped from the OneMap API, which takes in specific HDB address information and outputs the exact location of the HDB Block in (x,y) coordinates.
\item Population Dataset (https://worldpop.org/):
The data comes in a raster file that provides us with a pre-computed estimation of the population per pixel for the island of Singapore according to official United Nations population estimates.
\end{itemize}

\subsection{Hexagonal Grid Approach}

\begin{figure}[h!]
\includegraphics[width=0.45\textwidth]{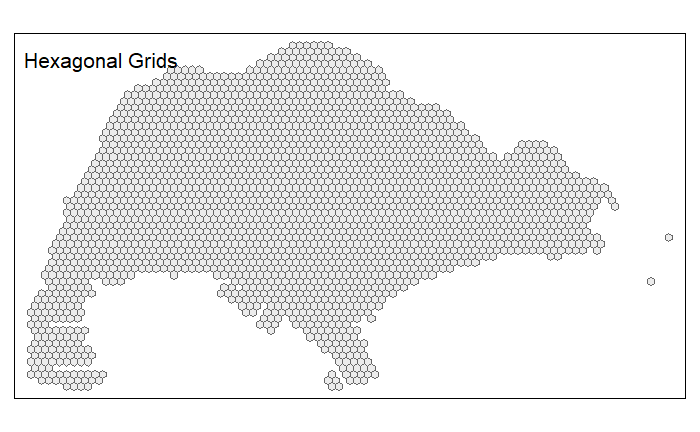}
\centering  
\caption{600m Hexagonal Grid}
\label{hexSize}
\end{figure}

As we are combining multiple different datasets (each with their own spatial resolution) for our analysis, we first utilize a common geospatial hexagonal grid structure of Singapore. By comparing the variables on the same hexagonal grid size, variables from different datasets will have a common basis for analysis. To study how urban features impact the UHI effect, there is a need to optimize the degree of granularity of the analysis; this is done by viewing all the data through different hexagon sizes as exemplified in Figure~\ref{hexSize}. 
The various methods employed will produce different outcomes with varying fits on the data for different hexagonal grid sizes, thereby producing models with different coefficient and error values. 
to address this potential issue, we decided empirically to use hexagon grid sizes of diameters 600m, 500m, 480m, 400m, 300m, and 200m to represent and compare the data. 

\subsection{Dependent Variable: Land Surface Temperature (LST)}
\begin{figure}[h!]
\includegraphics[width=0.5\textwidth]{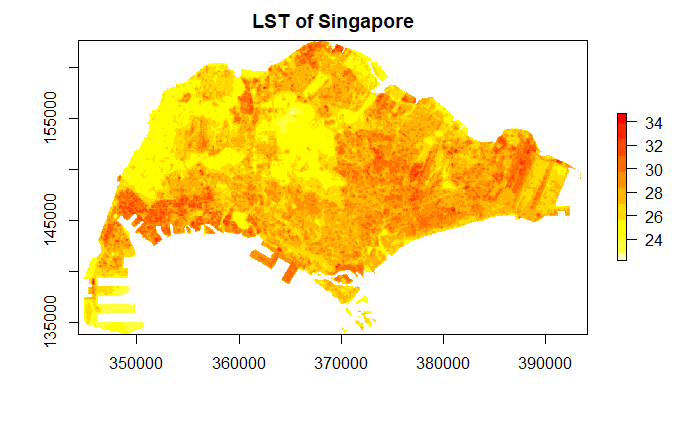}
\centering  
\caption{LST Raster Data from Landsat 8 at a per pixel level}
\label{LSTpixel}
\end{figure}

Figure~\ref{LSTpixel} shows the variation in LST (in degrees Celsius) for the island of Singapore on May 24, 2018. Hotter regions have a redder color while the cooler regions are towards the yellower end of the spectrum. LST is used as a computationally efficient measure of the UHI effect and the basic data unit is captured at the per pixel level. Using raster data from the Landsat 8 Satellite, the Land Surface Temperature (LST) variable can be computed using the Thermal Infrared Sensor (TIRS) Band 10 according to steps from the United States Geological Survey (USGS)~\cite{usgs2020} and Reddy et al.~\cite{reddy2017}. The data points are then represented at the hexagon level with the other variables. This variable can be obtained according to the following steps: 
\\
\subsubsection{Compute Top-of-Atmosphere (TOA) Radiance, L}
\begin{equation}
L = M_{L}Q_{cal} + A_{L}
\end{equation}
\(M_{L}\) : Band-specific multiplicative rescaling factor from the metadata\\
\(A_{L}\): Band-specific additive rescaling factor from the metadata  \\
\(Q_{cal}\):Quantized and calibrated standard product pixel values (DN)\\
\\
\subsubsection{Compute Brightness Temperature, T}
\begin{equation}
T = \frac{K_2}{ln(\frac{K_1}{L_\lambda} + 1)}
\end{equation}
\(L_\lambda\) = TOA spectral radiance (Watts/( m2 X srad X $\mu$ m))\\
\(K_1\) = Band-specific thermal conversion constant from the metadata \\
\(K_2\) = Band-specific thermal conversion constant from the metadata \\
\\
\subsubsection{Compute Proportion of Vegetation, Pv, using NDVI}
\begin{equation}
NDVI =\frac{(NIR - Red)}{(NIR + Red)}
\end{equation}
\begin{equation}
P_v =[\frac{NDVI - NDVI_{min}}{NDVI_{max} -  NDVI_{min}}]^2
\end{equation}
\\
\subsubsection{Compute Land Surface Emissivity, \(\epsilon\)}
\begin{equation}
 \epsilon=0.004P_v + 0.986
\end{equation}
\\
\subsubsection{Compute Land Surface Temperature (LST)} 
\begin{equation}
LST = \frac{T}{1 + [(\frac{\lambda T}{\rho})(ln\epsilon)]}
\end{equation}	
\(\lambda\) = average wavelength of band 10 \\
\(\rho\)= h x c/(\(\sigma\), which is equal to 1.438 x 10-2 mK , where \(\sigma\) is the
Boltzmann constant (1.8 x 10-23JK-1), h is Planck's constant
(6.626 x 10-34) and c is the velocity of light (3.0 x 108 ms-1)\\
 
{Finally, LST values are extracted into hexagonal grids through the mean LST value for each hexagon.}

\subsection{Independent Variables: Surface Information}
Similar to the LST dependent variable, the independent variables from this data category are highly granular as they are obtained from raster images at a per pixel level before they could be represented at the hexagon grid level. We describe a series of these independent variables that represent surface information and building information.

\begin{figure}[h!]
\includegraphics[width=0.5\textwidth]{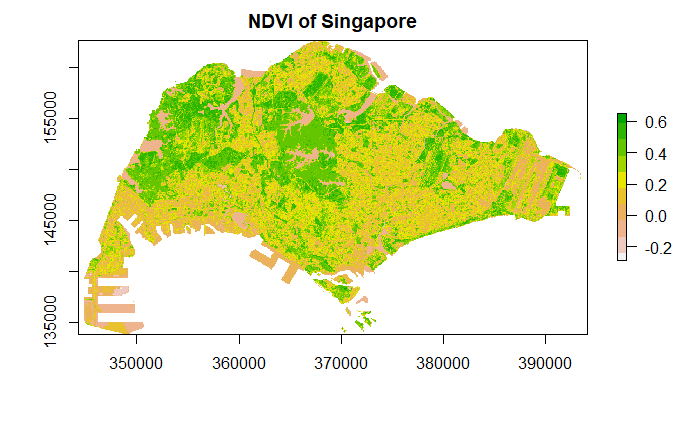}
\centering  
\caption{NDVI Raster Data from Landsat 8 at a per pixel level}
\label{ndviPixel}
\end{figure}

\subsubsection{Normalized Difference Vegetation Index (NDVI)}
Figure~\ref{ndviPixel} shows the variation in NDVI for the island of Singapore on May 24, 2018. Regions with higher concentrations of vegetation have a darker green color while the regions with lower concentrations of vegetation are towards the pinker end of the spectrum. NDVI measures the amount of vegetation cover in a given area via remote sensing method through the estimation of plant chlorophyll content. The study of this variable allows us to investigate if the UHI effect, as measured by LST, is influenced by the presence of vegetation and to understand the significance of the relationship. NDVI is calculated using the visible wavelength red band (Red: 0.636m - 0.674m) and Near Infra-Red band (NIR:  0.851m~0.879m) of the Landsat 8 Operational Land Imager (OLI) and Thermal Infrared Sensor (TIRS). The equation to compute NDVI is given as follows:

\begin{equation}
NDVI =\frac{(NIR - Red)}{(NIR + Red)}
\end{equation}

NDVI values are extracted into hexagonal grids through aggregation.

\begin{figure}[h!]
\includegraphics[width=0.5\textwidth]{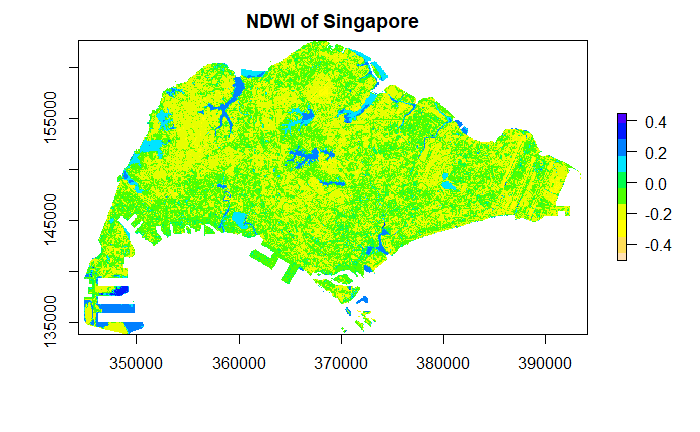}
\centering  
\caption{NDWI Raster Data from Landsat 8 at a per pixel level}
\label{ndwiPixel}
\end{figure}

\subsubsection{Normalized Difference Water Index (NDWI)}
Figure~\ref{ndwiPixel} shows the variation in NDWI for the island of Singapore on May 24, 2018. Regions with higher concentrations of water are more blue while the regions with lower concentrations of water are towards the more pinkish end of the spectrum. Similar to NDVI, NDWI measures the amount of moisture on land surface and in vegetation. The study of this variable allows us to investigate if the UHI effect (LST) is affected by the presence of water bodies and determine the significance of the relationship. It is computed using the visible wavelength green band (Green: 0.533m - 0.590m) and Shortwave Infra-Red band (SWIR: 1.566m~1.651m) of the Landsat 8 OLI and TIRS. The equation is as follows:

\begin{equation}
NDWI =\frac{(Green - SWIR)}{(Green + SWIR)}
\end{equation}

NDWI values are extracted into hexagonal grids through aggregation.

\begin{figure}[h!]
\includegraphics[width=0.5\textwidth]{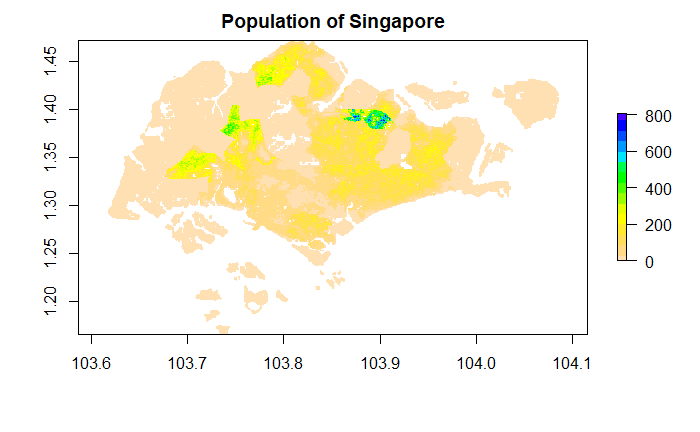}
\centering  
\caption{Population Raster Data from WorldPop at a per pixel level}
\label{populationPixel}
\end{figure}

\subsubsection{Total Population}
Figure~\ref{populationPixel} shows the variation in population numbers for the island of Singapore. Regions with higher concentrations of people are more blue while the regions with lower concentrations of people are towards the more pinkish end of the spectrum. This variable is an estimate of the total number of people per pixel that is computed using a mapping approach in the form of a Random Forest-based dasymetric redistribution (done by WorldPop). The study of this variable allows us to investigate if the UHI effect (LST) is affected by the presence and number of people in an area, and thereby determine the significance of the relationship. It is extracted into hexagonal grids through aggregation.

\subsection{Independent Variables: Building Information}
The variables from this data category are captured at the HDB Block level before they could be represented at the hexagon grid level. 

\subsubsection{Total Height}
This variable captures the sum of heights of every HDB block in terms of the number of storeys for each hexagonal grid. The study of this variable allows us to find out if the UHI effect (LST) is influenced by the height of residential HDB Blocks and determine the significance of the relationship. 

\subsubsection{Total Dwelling Units, Total Residential, Total Commercial, Total Market/Hawker, Total Multistorey Carpark, Total Precinct Pavilion, Total Miscellaneous}\footnote{Miscellaneous examples include admin office, childcare centre, education centre, Residents' Committees centre}
Some of Singapore's HDB Blocks are integrated developments where other amenities are built into the same floor area. These variables represent binary indicators for an amenity for each HDB block that are aggregated to give the total count of the binary presence of amenities for each hexagonal grid.  By investigating the relationship between these variables and the UHI effect (measured by LST), we can gain better insight into how statistically significant they could be in mitigating UHI.

\section{Statistical Modelling}
\label{sectFramework-2}
This study seeks to investigate the relationship between UHI, represented by LST, and the characteristics of Singapore’s urban form, represented by surface and building data, for different hexagonal grid sizes. For each of the hexagonal grid sizes, we use model selection criteria such as R-squared, Log-likelihood, Akaike Information Criterion (AIC), and the Moran I’s statistic to choose the most optimal model that describes the relationship between UHI and urban form characteristics. Between the two spatial models, a Lagrange Multiplier Test is applied to reveal which model better accounts for the spatial element in the data. We also compute and note the statistical significance of each variable and investigate how their status changes over the different hexagonal grid sizes and models. The regression models used are: Ordinary Least Squared (OLS), Spatial Lag Model (SAR), and Spatial Error Model (SEM). Variables are added incrementally and considered according to their statistical significance.

\subsection{OLS Regression}
\label{sectOLSRegression}
The OLS regression model is employed to explain the linear relationship between LST and the characteristics of Singapore’s urban form. In a similar study, OLS serves as an urban geocomputation technique used to quantify the relationship between a dependent variable and a list of geospatial factors~\cite{Voelkel2018}. The computed LST dependent variable is regressed against the independent variables consisting of surface and building information according to this equation: 

\begin{equation}
LST = \alpha + \sum_{i=1}^{n}{\beta}_i x_i + \epsilon
\end{equation}

where \(\alpha\) the intercept, \(\beta_i\) the regression coefficients, \(x_i\) the independent variables, and \(\epsilon\) the error term.

\subsection{Spatial Regression: Spatial Lag Model (SAR)}
\label{sectSpatialRegression}
In order to account for the potential spatial effects of UHI that could arise due to atmospheric flows and exchanges, we consider spatial regression models like SAR models~\cite{LeSagePace2009}. The SAR model is used to explain the relationship between LST and the elements of Singapore’s urban form with consideration for the spatial autocorrelation effects that the LST in one hexagon would have on its neighboring hexagons. The equation describing the relationship is as follows:

\begin{equation}
LST = \alpha + \sum_{i=1}^{n}{\beta}_i x_i + \rho W(LST) + \epsilon
\end{equation}

where \(\alpha\) is the intercept, \(\beta_i\) the regression coefficients, \(x_i\) the independent variables, \(\rho\) the spatial lag coefficient, \(W\) the spatial weights matrix, and \(\epsilon\) the error term. The sign (positive or negative) of the regression coefficient \(\beta_i\) determines the relationship that the independent variable \(x_i\) has with LST while its magnitude measures the influence it has on LST. The spatial lag \(\rho\) coefficient  captures the magnitude of the spatial lag present; a larger value indicates that the model can better account for spatial autocorrelation.

\subsection{Spatial Regression: Spatial Error Model (SEM)}
\label{sectSEM}
Similarly, the work done by LeSage and Pace~\cite{LeSagePace2009} explains how SEM models can account for spatial effects of UHI as well. The SEM model is used to explain the relationship between LST and the elements of Singapore’s urban form with consideration for the spatial autocorrelation effects that the residuals in one hexagon would have on its neighboring hexagons due to an unknown spatial covariate. The equation describing the relationship is as follows:

\begin{equation}
LST = \alpha + \sum_{i=1}^{n}{\beta}_i x_i + \lambda WU + \epsilon
\end{equation}

where \(\alpha\) is the intercept, \(\beta_i\) the regression coefficients, \(x_i\) the independent variables, \(\lambda\) the spatial error coefficient, \(W\) the spatial weights matrix, \(U\) the unspecified random error, and \(\epsilon\) the error term. The sign (positive or negative) of the regression coefficient \(\beta_i\) determines the relationship that the independent variable \(x_i\) has with LST while its magnitude measures the influence it has on LST. The spatial error coefficient \(\lambda\) captures the magnitude of the spatial error present; a larger value indicates that the model can better account for spatial autocorrelation.

\section {Experimental Results \& Discussions}
\label{sectResults}
This section looks at the findings and insights we have obtained from our Statistical Modelling 
experiments that attempt to answer our research questions on the Urban Heat Island (UHI) effect, namely:

\begin{itemize}
    \item RQ 1: How is the UHI effect influenced by the urban form?
    \item RQ 2: Which elements of the urban form are worth investigating?
\end{itemize}

Table 2 contains the experimental results for the Statistical Modelling section. 
For our research paper, the urban form will comprise of city-level elements found in Table 2, such as the Normalized Difference Vegetation Index (quantifies amount of green spaces in Singapore) and Total Height (quantifies heights of HDB blocks) belonging to two categories: surface information and building information. 

\subsection{Statistical Modelling Findings}

\begin{table*}[h!]
\centering  
\includegraphics[width=0.495\textwidth]{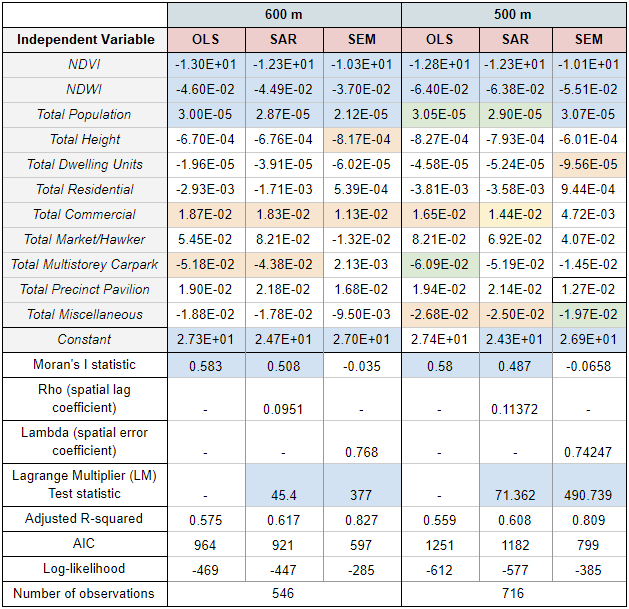}
\includegraphics[width=0.495\textwidth]{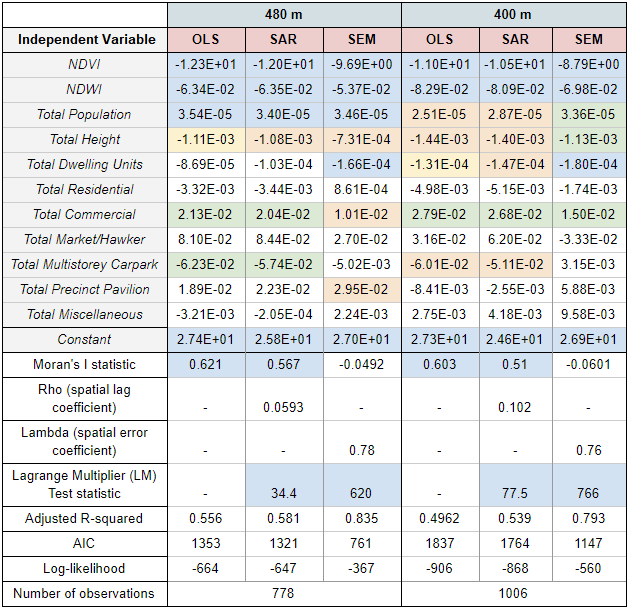}
\includegraphics[width=0.995\textwidth]{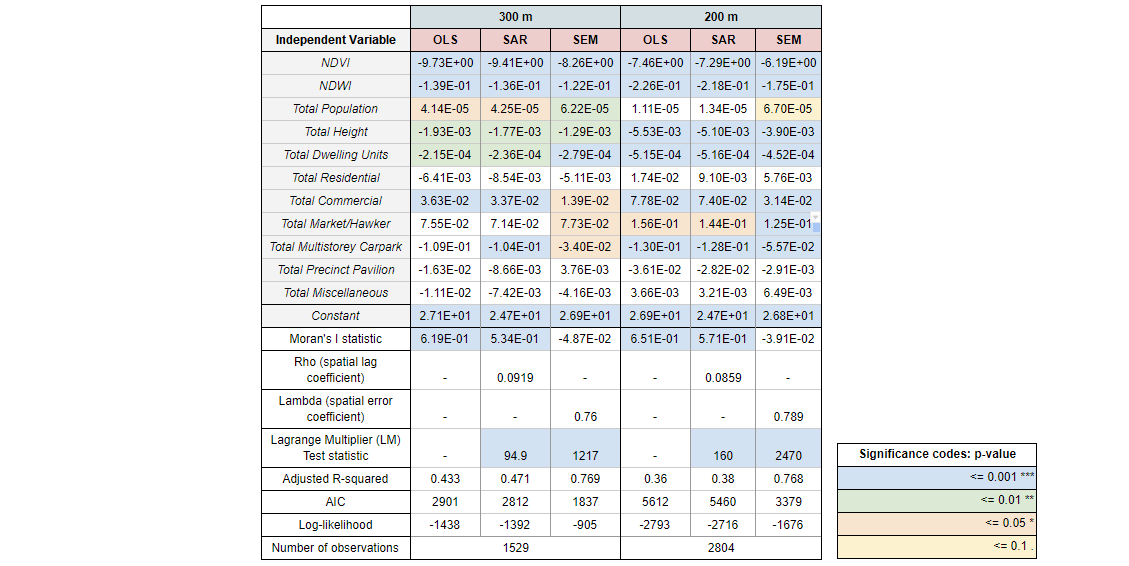}
\caption{Statistical Modelling Results}
\label{statModelResults}
\end{table*}

% \bigskip
%\bigskip
The statistical modelling process considers three different model types (OLS, SAR, and SEM) across different hexagon sizes of 600m, 500m, 480m, 400m, 300m, and 200m as shown in Table~\ref{statModelResults}. The hexagon sizes were decided empirically to give a reasonable range of different results for comparison. 

\subsubsection{Model Selection}
\label{sectModelSelection}
Comparing across different hexagon sizes, the relatively larger hexagon grid structures of 600m, 500m, and 480m yielded better performing statistical models in terms of the model selection criteria such as adjusted r2 , AIC, Log-likelihood. For example, with regards to Table~\ref{statModelResults}, r2 stayed above 0.550 for OLS models for the 600m , 500m, 480m models but fell below 0.500 for the 400m , 300m, and 200m models. Similarly, AIC values remained below 1000 for the 600m, 500m, 480m SEM models but entered the 1000s range for the smaller hexagon grid structures of 400m,  300m, and 200m SEM models.

Comparing across the three different types of statistical models, SEM models performed consistently (across all hexagon grid sizes) and significantly better than SAR and OLS models with OLS models having the poorest performance. For instance, with regards to Table~\ref{statModelResults}, OLS models have the r2 in the lowest range of 0.360 - 0.575, SAR models have r2 in the mediocre range of 0.380 - 0.617, while SEM models have r2 values in the highest range of 0.768 - 0.835. Moreover, the Moran’s I Statistic is very significant for the OLS and SAR models at 99.9\% significance levels but not so for SEM models, indicating that the autocorrelation in the dataset can best be accounted for by SEM models. Comparing between the two types of spatial models (SAR and SEM), the magnitude of SEM’s LM Test Statistic (range: 377 - 2470) always largely outweighs that of SAR’s (range: 34.4 - 160) despite both being significantly non-zero. SEMs account for spatial dependency in the errors (residuals) of the model which are caused by an unmeasured spatial covariate. Regardless of the hexagon size, the SEM models outperform both OLS and SAR models.

Our experimental results show the best performing model to be the SEM model at a hexagon grid structure of 480m. With regards to Table~\ref{statModelResults}, the model best explains data variability with an adjusted r2 of 0.835 with the next best model being the 600m SEM model having an adjusted r2 of 0.827. However, our best performing 480m SEM model’s fit to the data (AIC = 761, Log-likelihood = -367) is second to the 600m SEM model’s fit (AIC = 597, Log-likelihood = -285). Nonetheless, it captures more significant relationships between our variables and the UHI effect (measured by LST). Moreover, the spatial dimension is better accounted for according to the higher value of the spatial error coefficient, , which is 0.780 for the 480m SEM model but 0.768 for the 600m SEM model. Hence, we selected the 480m SEM model to be our best performing model. 

\subsubsection{Statistically Significant Variables}
The variables in this section are statistically significant for the majority of all models considered and especially so for our best performing 480m SEM model. The existence of some discrepancies in changing levels of significance can provide further indication on model suitability and variable importance.

\textbf{NDVI \& NDWI}:  
LST decreases when NDVI and NDWI increase. For the 480m SEM model, NDVI = -9.69E+00 while NDWI = -5.37E-02 according to Table~\ref{statModelResults}. The Urban Heat Island effect (measured by LST), is inversely proportional to NDVI and NDWI for all hexagon sizes at the highest level of significance of 0.999 (or 99.9\%). This suggests that the presence of vegetation has the effect of lowering temperatures due to evaporative cooling of the moisture content in the soil~\cite{Oke2017}. In addition, vegetation reflects thermal energy into the atmosphere~\cite{doick2013} while providing canopy cover that reduces incident radiation and the micro greenhouse effect~\cite{emmanuel2005,anyanwu2006}. Apart from lowering LST, green spaces have been shown to provide various health and emotional benefits to its residents~\cite{lim-smartcity19,wang-www18,lim-www18}. Similarly, the presence of water bodies nearby will reduce daytime land surface temperatures as well~\cite{Wu2019}. 
Therefore, the presence of water bodies and green spaces are pivotal in efforts to reduce UHI effects.

\textbf{Total Population}: 
The UHI (LST) effect is amplified by an increase in population. Using the best performing 480 SEM model as an example, we see from Table~\ref{statModelResults} that the variable has a positive coefficient of 3.46E-05 that is statistically significant at 99.9\%. The positive Total Population coefficient suggests that the existence of anthropogenic heat, generated by the presence of people and their activities, contributes to rising surface temperatures~\cite{shahmohamadi2011}. Improvements in the level of significance of this variable under the SEM model (as compared to OLS and SAR models) further implies the existence of spatial autocorrelation in its residuals due to unmeasured covariates that was not captured by OLS and SAR models.  Hence, we have a reinforced understanding that the presence of people contributes to the UHI effect with a spatial overflow to neighboring areas. 

\textbf{Total Height}:
The height of HDB blocks have a negative relationship with the UHI effect (LST). Using the best performing 480m SEM model as an example, we see from Table~\ref{statModelResults} that the variable has a negative coefficient of 7.31E-04 that is statistically significant at 95\%. Interestingly, it would appear that the presence of taller HDB blocks lowers the surrounding surface temperature. This contradicts what we expect of tall buildings to create thermal energy build-up and poor air circulation, which then increase the UHI effect~\cite{Oke1981,Eliasson1996,BottyanUnger2003}. A possible explanation for this negative relationship could be that taller HDB blocks provide more shade and shelter from the sun at different angles throughout the day, thereby reducing the amount of heat exposure of the surrounding surfaces in the area~\cite{Chen2020}. In addition, cross-sectional information is not accounted for in our dataset; well-designed HDB blocks with good airflow would be able to counter the expected effects of poor air circulation and thermal energy build-up~\cite{Low2014}. According to the Urban Redevelopment Authority (URA), the Singaporean government uses urban environmental modelling techniques to design and position HDB flats in a manner that maximises air flow and shade~\cite{URA2018}. Similar to the Total Population variable, improvements in the level of significance of this variable under the SEM model as compared to OLS and SAR further implies the existence of spatial autocorrelation in its residuals due to unmeasured covariates that was not captured by OLS and SAR models. Hence, we have a better understanding that the height of HDB blocks reduces the surrounding UHI effect not just within the hexagon unit, but also neighboring hexagons units. 

\textbf{Total Commercial}:
Across the majority of the models, this statistically significant variable has a positive influence on the UHI effect that suggests an increase in surface temperatures in areas where commercial facilities are present. Using the best performing 480m SEM model as an example, we see from Table~\ref{statModelResults} that the variable has a positive coefficient of 1.01E-02 that is statistically significant at 95\%. This could be attributed to the operational nature of such outlets (provision shops, healthcare facilities, F\&B etc.) where relatively higher electrical consumption and carbon output are expected. Furthermore, the relative statistical significance of this variable increases as the hexagon grids get smaller since their influence within the area would carry more weight in smaller grids. 

\textbf{Total Dwelling units}:
Results indicate that the UHI effect is inversely proportional to the total number of dwelling units in a HDB block, suggesting that the UHI effect is in fact lower in areas with more dwelling units. Using the best performing 480m SEM model as an example, we see from Table~\ref{statModelResults} that the variable has a negative coefficient of 1.66E-04 that is statistically significant at 99.9\%. At first glance, this may seem contradictory as we would expect an increase in surface temperatures in areas with more human dwelling activity. However, considering the time of day that the surface temperature data was captured via Remote Sensing (daytime on Thursday, 24 May 2018), it would be plausible for surface temperatures to be lower in HDB estates when compared to other parts of Singapore where industrial activities are taking place during the work day.  According to the National Environmental Agency (NEA), industrial areas are to be located away from residential areas; sufficient buffer spaces must be provided to keep residential developments away from industrial ones~\cite{NEA2019}. By comparison, hexagons containing residential units produce less thermal energy and so they would have a negative effect on LST. Hence, we can expect hexagons with residential buildings to reduce the UHI effect as they are less likely to contain industrial developments.

\textbf{Total Multistorey Carpark}:
Across a majority of models, this statistically significant variable has a negative relationship with the UHI effect (LST), indicating that an increase in the total number of multi-storey carparks resulted in lowered surface temperatures. Using the 300m SEM model as an example, we see from Table~\ref{statModelResults} that the variable has a negative coefficient of 3.40E-02 that is statistically significant at 95\%.This could be due to the way in which Singapore designs its multi-storey carparks with a rooftop garden on the highest floor~\cite{Chan2013}. From our previous insights on the NDVI and NDWI variables, we know that the presence of vegetation lowers surface temperatures through evaporative cooling of soil moisture. Moreover, existing research on using green roofs to combat UHI corroborates our findings by suggesting that green roofs “can be 30-40$^\circ$F lower than those of conventional roofs and can reduce city-wide ambient temperatures by up to 5$^\circ$F”~\cite{EPA2019}. Hence, our findings indicate that we could potentially reduce the UHI effect by modifying or retrofitting rooftops with green features. Consequently, it also posits the possibility of lowering the surface temperature of other impervious surfaces by applying either a horizontal or vertical green layer.

\section{Conclusion \& Future Work}
\label{sectConclusion}

This research has significantly improved our understanding of how urban form characteristics play a key role in influencing the Urban Heat Island phenomenon. Using remote sensing data acquired from satellite imagery, we are able to perform geospatial analytics to elucidate the relationship between urban form characteristics and the Urban Heat Island effect.
The statistical modelling aspect of the research reveals to us the importance of different urban form characteristics with specific regard to how they affect UHI. The presence of vegetation and water bodies play a tremendous role in tackling UHI; steps must be taken to infuse the urban landscape with more green and water features to reduce temperatures while careful land use planning can go a long way in reducing thermal buildups in our residential estates. Tall buildings, while initially believed to trap urban heat, can in fact improve thermal comfort by providing shade to the surrounding area and if cleverly ventilated with good airflow, they can potentially mitigate heat storage.  

Nonetheless, this research does have its limitations. The analysis is dependent on an overall 2-dimensional dataset. Given sufficient resources to procure 3-dimensional data or adequate access to such existing data, the results of this research could potentially yield much more insight. Moreover, the building data is limited to residential HDB blocks. The availability of additional building data including commercial and industrial facilities would definitely improve our findings. Along this direction, future work can further expand this research to include different times of the day across multiple time periods for more enriching results.

There are other extensions to this work that serve as future research directions. For example, one can develop machine learning models and systems to predict the heat levels in different parts of a country, based on multi-modal data from different sources~\cite{tonekaboni2018scouts}. In addition to sensor and map-based data, one can enhance such measurements and predictions with crowd-sourced data from social media~\cite{Heng-BigData20}, mobile phones~\cite{tonekaboni2018scouts} and other relevant human sensors.

\section*{Acknowledgment}
This research is funded in part by the Singapore University of Technology and Design under grant SRG-ISTD-2018-140. The authors would like to thank Olivia Nicol for her useful suggestions.



\bibliographystyle{IEEEtran}
\balance
\bibliography{urbanHeat}

\end{document}